# Emergent elasticity of skyrmion crystal in chiral magnets: the constitutive equations

Yangfan Hu[1,*], Biao Wang[1]

[1]Sino-French Institute of Nuclear Engineering and Technology, Sun Yat-Sen University, 510275 GZ, China

**Abstract**
Emergent phenomena refer to presence of novel elementary structure with a larger length scale due to collective cooperation between the components underneath. A general challenge to understand diverse kinds of emergent phenomena, including the origin of life[1], the behavior of large networks[2], and many novel states appearing in condensed matter[3,4], is to establish a mathematic formula that connects them with the properties of the components underneath. Recent observation of the skyrmion crystal (skX) appearing in helimagnets[5-13] provides a two-fold emergent phenomenon: first the emergence of magnetic skyrmion, a two-dimensional spin whirl with topological properties [14-16], then the crystallization of skyrmions into skX[17]. Compared with a localized skyrmion, skX is a novel magnetic state with emergent macroscopic properties. To begin with, it deforms in FeGe when uniaxial tension is applied, with an elastic stiffness two orders of magnitude smaller than that of the underlying material[8]. The emergent elastic property makes it convenient to manipulate the anisotropy of texture, mobility, and other properties of skX by applying mechanical forces, while the physical mechanism behind is not yet understood. Here we derive the linear constitutive equations governing the emergent elastic behavior of skX under applied mechanical forces as well as other effective fields. The linear coefficients appearing in the equations are expressed analytically in terms of the thermodynamic parameters describing the magnetoelastic interaction of the underlying chiral magnet. Through the deduction of equations, we clarify that two types of coupled deformation may occur for skX, that is the internal deformation, described by variation of the Fourier magnitudes, and the lattice deformation, described by the emergent elastic strains. We also find that besides mechanical loads, bias magnetic field, spatially periodic magnetic field, spatially periodic mechanical loads, and emergent stresses can all lead to deformation of the skX. When homogeneous stresses are applied only, the emergent elastic strains are linearly related to the elastic strains through the *emergent strain ratio matrix* $\boldsymbol{\lambda}$. We calculate $\boldsymbol{\lambda}$ for the skX in bulk MnSi and discuss its variation with the temperature and magnetic field. In particular, we find that all components of $\boldsymbol{\lambda}$ are sensitive to variation of magnetic field $b$, such that they change sign as $b$ increases. Moreover, as $b$ approaches the critical magnetic field of the skX-ferromagnetic phase transition, the dominant components of $\boldsymbol{\lambda}$ diverges. The constitutive equations are of fundamental significance in emergent elasticity, a new area studying the variation of internal forces and deformation of emergent crystals and their relation when exposed to effective external fields.

## I. Introduction

Observation of both skX and bi-skyrmion crystal in various kinds of bulk magnets[5,8,9] and corresponding thin film structures[6,7,10-13,18,19] suggests that crystallization of emergent particles is by no means fortuitous. Firstly, topological emergent particles may exist in many different systems with chiral interactions, such as B20 helimagnets, non-centrosymmetric ferroelectrics, and chiral liquid crystals. Secondly, when a type of emergent particle becomes locally stable, there is a general tendency for it to occupy the whole space so as to achieve a globally stable state. Finally, as a result of competition between the efficiency to occupy the whole space and the necessity to achieve the lowest possible local free energy of a single emergent particle, an emergent crystalline state with certain type of symmetry should appear. Due to the general incompatibility between the global crystalline symmetry and the local axial symmetry of the isolated emergent particle concerned, there is usually a local symmetry breaking for the emergent particles during the crystallization[20]. For this, the emergent crystals derive from but go beyond isolated emergent particles. Studying their macroscopic properties not only provides us with a deeper understanding of this new state of matter, but will also stimulate novel technological applications.

This work is enlightened by the uniaxial tension experiment of FeGe, where the skX inside is found to undergo significant deformation two orders of magnitude larger than the deformation of the underlying atomic lattice[8]. Does skX always deform under application of mechanical loads? If yes, what properties of the underlying material determine such deformation? What other external fields also lead to deformation of skX? Why skX deforms the way it does when exposed to mechanical forces? In this work, we derive the linear constitutive equations for the emergent elastic deformation of skX in chiral magnets due to application of mechanical forces as well as other effective fields. The equations are derived based on an extended Landau-Ginzburg functional of the underlying helimagnet incorporating a comprehensive description of magnetoelastic interaction[21]. By doing so, we established an explicit relation between the emergent elastic property of skX and the magnetoelastic property of the underlying material, and provide answers or clues to the questions raised above. The methodology constructed in this paper is generally applicable to discussion of any emergent macroscopic property of any emergent crystal, once you have obtained a reliable free energy density functional of the underlying system.

**II. Formulation**

The primary issue is how to describe the deformation of the skX mathematically. We study the question within the following rescaled Landau-Ginzburg functional incorporating magnetoelastic coupling[5,21-24] (the rescaling process is shown in ref. [21]).

$$\widetilde{w}(\mathbf{m}, \varepsilon_{ij}, \gamma_{ij}) = \sum_{i=1}^{3} \left(\frac{\partial \mathbf{m}}{\partial r_i}\right)^2 + 2\mathbf{m} \cdot (\nabla \times \mathbf{m}) - 2\mathbf{b} \cdot \mathbf{m} + t\mathbf{m}^2 + \mathbf{m}^4 + \widetilde{w}_{el} + \widetilde{w}_{me}, \quad (1)$$

where

$$\widetilde{w}_{el} = \frac{1}{2}\tilde{C}_{11}(\varepsilon_{11}^2 + \varepsilon_{22}^2 + \varepsilon_{33}^2) + \tilde{C}_{12}(\varepsilon_{11}\varepsilon_{22} + \varepsilon_{11}\varepsilon_{33} + \varepsilon_{22}\varepsilon_{33})$$
$$+ \frac{1}{2}\tilde{C}_{44}(\gamma_{12}^2 + \gamma_{13}^2 + \gamma_{23}^2) \quad (2)$$

denotes the elastic energy density of materials with cubic symmetry, and

$$\widetilde{w}_{me} = \tilde{L}_1(m_1^2 \varepsilon_{11} + m_2^2 \varepsilon_{22} + m_3^2 \varepsilon_{33}) + \tilde{L}_2(m_3^2 \varepsilon_{11} + m_1^2 \varepsilon_{22} + m_2^2 \varepsilon_{33})$$
$$+ \tilde{L}_3(m_1 m_2 \gamma_{12} + m_1 m_3 \gamma_{13} + m_2 m_3 \gamma_{23}) + \widetilde{K}\mathbf{m}^2 \varepsilon_{ii} + \sum_{i=1}^{6} \tilde{L}_{Oi} \tilde{f}_{Oi}, \quad (3)$$

denotes the magnetoelastic energy density, where

$$\begin{aligned}
\tilde{f}_{O1} &= \varepsilon_{11}(m_{1,2}m_3 - m_{1,3}m_2) + \varepsilon_{22}(m_{2,3}m_1 - m_{2,1}m_3) + \varepsilon_{33}(m_{3,1}m_2 - m_{3,2}m_1), \\
\tilde{f}_{O2} &= \varepsilon_{11}(m_{3,1}m_2 - m_{2,1}m_3) + \varepsilon_{22}(m_{1,2}m_3 - m_{3,2}m_1) + \varepsilon_{33}(m_{2,3}m_1 - m_{1,3}m_2), \\
\tilde{f}_{O3} &= \varepsilon_{11}m_1(m_{2,3} - m_{3,2}) + \varepsilon_{22}m_2(m_{3,1} - m_{1,3}) + \varepsilon_{33}m_3(m_{1,2} - m_{2,1}), \\
\tilde{f}_{O4} &= \gamma_{23}(m_{1,3}m_3 - m_{1,2}m_2) + \gamma_{13}(m_{2,1}m_1 - m_{2,3}m_3) + \gamma_{12}(m_{3,2}m_2 - m_{3,1}m_1), \\
\tilde{f}_{O5} &= \gamma_{23}(m_{3,1}m_3 - m_{2,1}m_2) + \gamma_{13}(m_{1,2}m_1 - m_{3,2}m_3) + \gamma_{12}(m_{2,3}m_2 - m_{1,3}m_1), \\
\tilde{f}_{O6} &= \gamma_{23}m_1(m_{3,3} - m_{2,2}) + \gamma_{13}m_2(m_{1,1} - m_{3,3}) + \gamma_{12}m_3(m_{2,2} - m_{1,1}).
\end{aligned} \quad (4)$$

The skX phase appears as a periodic solution of rescaled magnetization[20]:

$$\mathbf{m} = \mathbf{m_0} + \sum_{i=1}^{\infty} \sum_{j=1}^{n_i} \mathbf{m}_{\mathbf{q}_{ij}} e^{i\mathbf{q}_{ij}\cdot\mathbf{r}}, \quad (5)$$

where $\mathbf{q}_{11}$, $\mathbf{q}_{12}$, and $\mathbf{q}_{13}$ form an equilateral triangle in the $x$-$y$ plane when connected head-to-tail for hexagonal skX. The other $\mathbf{q}_{ij}$ satisfy $\mathbf{q}_{ij} = k\mathbf{q}_{11} + l\mathbf{q}_{12}$, where $k, l$ are integers. For any subscript $i$ we have $|\mathbf{q}_{i1}| = |\mathbf{q}_{i2}| = |\mathbf{q}_{i3}| = \cdots |\mathbf{q}_{in_i}|$, and for any subscript $j$ we have $|\mathbf{q}_{1j}| < |\mathbf{q}_{2j}| < |\mathbf{q}_{3j}| < \cdots$. Hence the skX can be regarded as a two-dimensional crystalline state with reciprocal vectors $\mathbf{q}_{11}$ and $\mathbf{q}_{12}$. Similar to the atomic lattice[25], the structure of the skyrmion lattice always guarantees the orthogonality between the lattice vectors and the reciprocal vectors:

$$\begin{aligned}
\mathbf{a}_1 \cdot \mathbf{q}_{11} &= 1, \\
\mathbf{a}_1 \cdot \mathbf{q}_{12} &= 0, \\
\mathbf{a}_2 \cdot \mathbf{q}_{11} &= 0, \\
\mathbf{a}_2 \cdot \mathbf{q}_{12} &= 1,
\end{aligned} \quad (6)$$

where $\mathbf{a}_1$, $\mathbf{a}_2$ denote the lattice vectors. When skX deforms, the original lattice vectors $\mathbf{a}_1^u$, $\mathbf{a}_2^u$ transform to the deformed lattice vectors $\mathbf{a}_1^d$, $\mathbf{a}_2^d$. Based on the Cauchy-Born law[25], we introduce the ***emergent elastic strains*** (EES) $\varepsilon_{\alpha\beta}^e, (\alpha, \beta = 1,2)$, and the ***emergent rotation angle (ERA)*** $\omega^e$, which describe the deformation of skX through

$$\mathbf{a}_\alpha^d = \mathbf{F}\,\mathbf{a}_\alpha^u, \quad (\alpha = 1, 2), \quad (7)$$

where $\mathbf{F} = \begin{bmatrix} 1 + \varepsilon_{11}^e & \varepsilon_{12}^e + \omega^e \\ \varepsilon_{12}^e - \omega^e & 1 + \varepsilon_{22}^e \end{bmatrix}$, and the EES and ERA are related to the ***emergent displacement field*** $u_\alpha^e, (\alpha = 1,2)$ by $\varepsilon_{\alpha\beta}^e = \frac{1}{2}(u_{\alpha,\beta}^e + u_{\beta,\alpha}^e)$, $(\alpha, \beta = 1,2)$, and $\omega^e = \frac{1}{2}(u_{1,2}^e - u_{2,1}^e)$. The deformed wave vectors $\mathbf{q}_{11}^d$ and $\mathbf{q}_{12}^d$ as functions of $\varepsilon_{\alpha\beta}^e$ and $\omega^e$ can be solved by substituting eq. (3) into eq. (2), and $\mathbf{q}_{13}^d$ is derived from $\mathbf{q}_{11}^d + \mathbf{q}_{12}^d + \mathbf{q}_{13}^d = 0$. Assume $\mathbf{q}_{11}^u = q[0,1]^T$, $\mathbf{q}_{12}^u = q[-\frac{\sqrt{3}}{2}, -\frac{1}{2}]^T$, we have after manipulation

$$\begin{aligned}
\mathbf{q}_{11}^d &= q\left[\frac{\omega^e - \varepsilon_{12}^e}{1 + \varepsilon_{11}^e + \varepsilon_{22}^e + \varepsilon_{11}^e\varepsilon_{22}^e - (\varepsilon_{12}^e)^2 + (\omega^e)^2}, \frac{1 + \varepsilon_{11}^e}{1 + \varepsilon_{11}^e + \varepsilon_{22}^e + \varepsilon_{11}^e\varepsilon_{22}^e - (\varepsilon_{12}^e)^2 + (\omega^e)^2}\right]^T, \\
\mathbf{q}_{12}^d &= q\left[\frac{-\sqrt{3} - \sqrt{3}\varepsilon_{22}^e + \varepsilon_{12}^e - \omega^e}{2[1 + \varepsilon_{11}^e + \varepsilon_{22}^e + \varepsilon_{11}^e\varepsilon_{22}^e - (\varepsilon_{12}^e)^2 + (\omega^e)^2]}, \frac{-1 - \varepsilon_{11}^e + \sqrt{3}(\varepsilon_{12}^e + \omega^e)}{2[1 + \varepsilon_{11}^e + \varepsilon_{22}^e + \varepsilon_{11}^e\varepsilon_{22}^e - (\varepsilon_{12}^e)^2 + (\omega^e)^2]}\right]^T.
\end{aligned} \quad (8)$$

Substituting eq. (8) into eq. (5), we have for the deformed skX $\mathbf{m} = \mathbf{m}(\mathbf{m_c}, \mathbf{m}_{\mathbf{q}_{ij}}, \varepsilon_{\alpha\beta}^e, \omega^e)$, where $q$ in eq. (8) is the length of wave vectors before the deformation, and is thus considered a known parameter. The deformation of the skX can be categorized into two types: internal deformation (induced by variation of $\mathbf{m}_{\mathbf{q}_{ij}}$) and lattice deformation (induced by variation of $\varepsilon_{\alpha\beta}^e$ or $\omega^e$). As illustrated in Figure 1(a-d), internal deformation and lattice deformation both lead to changes of the magnetization distribution pattern, while only lattice deformation changes the skX lattice constant.

In practice, the equilibrium values of $\mathbf{m}_{\mathbf{q}_{ij}}$ and $\varepsilon^e_{\alpha\beta}$, $\omega^e$ are coupled, so that lattice deformation is usually accompanied by internal deformation and vise versa. $\varepsilon^e_{11}$, $\varepsilon^e_{22}$, $\varepsilon^e_{12}$ and $\omega^e$ correspond to four basic emergent deformation modes of skX, which are elongation in the $x$ and $y$ direction, shear deformation and "emergent rotation", i.e., rigid rotation of the skX with respect to the underlying atomic lattice. The deformed Wigner-Seitz cell for skX for the four basic modes is illustrated in Figure 1(e-h).

To know what kind of external fields can cause emergent deformation of skX, we first decide all the independent variables permitted by the thermodynamic model introduced in eq. (1). Presence of skX in the material leads to appearance of periodic elastic field[26,27], for which the total elastic strains can be expanded as

$$\varepsilon_{ij}(r) = \bar{\varepsilon}_{ij} + \frac{1}{2}\sum_{\mathbf{q}\neq \mathbf{0}}\left[-\left(U_i^{qc}q_j + U_j^{qc}q_i\right)\sin \mathbf{q}\cdot\mathbf{r} + \left(U_i^{qs}q_j + U_j^{qs}q_i\right)\cos \mathbf{q}\cdot\mathbf{r}\right], \qquad (9)$$

which are composed of a homogeneous part $\bar{\varepsilon}_{ij}$ and a periodic part induced by a periodic displacement field $\sum_{\mathbf{q}\neq\mathbf{0}}(\mathbf{U}^{qc}\cos\mathbf{q}\cdot\mathbf{r} + \mathbf{U}^{qs}\sin\mathbf{q}\cdot\mathbf{r})$. The expansion is only performed for those $\mathbf{q}$ which have a coupling with the Fourier components of the skX. From eqs. (5, 8, 9), we list all independent variables into four vectors as follow

$$\boldsymbol{\varepsilon}^e = [\varepsilon^e_{11}, \varepsilon^e_{22}, \varepsilon^e_{12}, \omega^e]^T, \qquad (10)$$

$$\bar{\boldsymbol{\varepsilon}} = [\bar{\varepsilon}_{11}, \bar{\varepsilon}_{22}, \bar{\varepsilon}_{33}, \bar{\gamma}_{23}, \bar{\gamma}_{13}, \bar{\gamma}_{12}]^T, \qquad (11)$$

$$\mathbf{U}^{\mathbf{q}} = [U_1^{q_{11}c}, U_2^{q_{11}c}, U_3^{q_{11}c}, U_1^{q_{11}s}, U_2^{q_{11}s}, U_3^{q_{11}s}, U_1^{q_{12}c}, \ldots, U_3^{q_{12}s}, U_1^{q_{13}c}, \ldots, U_3^{q_{13}s}, U_1^{q_{21}c}, \ldots]^T, (12)$$

$$\mathbf{m}^{\mathbf{a}} = [m_{01}, m_{02}, m_{03}, c^{re}_{111}, c^{re}_{121}, c^{re}_{131}, c^{re}_{112}, \ldots, c^{re}_{133}, c^{im}_{111}, c^{im}_{121}, c^{im}_{131}, c^{im}_{112}, \ldots, c^{im}_{133}, c^{re}_{211}, \ldots]^T, (13)$$

Where the relation between $c^{re}_{ijk}$, $c^{im}_{ijk}$ and $\mathbf{m}_{\mathbf{q}_{ij}}$ is introduced in the method section. The length of $\mathbf{U}^{\mathbf{q}}$ in eq. (11) and $\mathbf{m}^{\mathbf{a}}$ in eq. (13) depends on the order of Fourier representation in eq. (1). For the $n$th order Fourier representation ($n \leq 3$), the length of $\mathbf{U}^{\mathbf{q}}$ is $18n$ and the length of $\mathbf{m}^{\mathbf{q}}$ is $18n + 3$.

The effective external fields to deform skX are described as the work conjugates of all the independent variables. We define the work conjugates of $\varepsilon^e_{11}, \varepsilon^e_{22}, \varepsilon^e_{12}$ as the emergent elastic stresses $\sigma^e_{11}, \sigma^e_{22}, \sigma^e_{12}$, and the work conjugate of $\omega^e$ as the emergent torque $\mathfrak{T}^e$. The work conjugates of the other independent variables can be derived as follow: the elastic strains $\bar{\varepsilon}_{11}, \bar{\varepsilon}_{22}, \bar{\varepsilon}_{33}, \bar{\gamma}_{23}, \bar{\gamma}_{13}, \bar{\gamma}_{12}$ versus the elastic stresses $\bar{\sigma}_{11}, \bar{\sigma}_{22}, \bar{\sigma}_{33}, \bar{\tau}_{23}, \bar{\tau}_{13}, \bar{\tau}_{12}$; $U_k^{\mathbf{q}c}$ versus $-\frac{1}{2}\sigma^{\mathbf{q}s}_{kl}q_l$; $U_k^{\mathbf{q}s}$ versus $\frac{1}{2}\sigma^{\mathbf{q}c}_{kl}q_l$, where $\sigma^{\mathbf{q}s}_{kl}$ and $\sigma^{\mathbf{q}s}_{kl}$ denote amplitude of spatially periodic stress field $\sigma^{\mathbf{q}c}_{kl}\cos\mathbf{q}\cdot\mathbf{r} + \sigma^{\mathbf{q}s}_{kl}\sin\mathbf{q}\cdot\mathbf{r}$; the first three components of $\mathbf{m}^{\mathbf{a}}$ versus the bias magnetic field $\mathbf{b}$; and the other components of $\mathbf{m}^{\mathbf{a}}$ versus $\mathbf{b}^*_{\mathbf{q}_{ij}}$, where $\mathbf{b}^*_{\mathbf{q}_{ij}}$ denotes amplitude of spatially periodic magnetic field $\mathbf{b}^*_{\mathbf{q}_{ij}}e^{-i\mathbf{q}_{ij}\cdot\mathbf{r}}$. We find five types of external fields that cause emergent deformation of skX, which are homogeneous mechanical forces, bias magnetic field, spatially periodic mechanical forces, spatially periodic magnetic field, and the emergent elastic stresses. From dimensional analysis, we see that emergent elastic stresses share the same dimension with the elastic stresses. Microscopically, the elastic stresses stem from atomic interaction while the emergent elastic stresses stem from skyrmion-skyrmion interaction.

Assume that the equilibrium state of the system is disturbed by a small variation of external fields, and the disturbance is small enough so that the state of the system changes smoothly, the linear constitutive equation which determines the emergent deformation of skX can be derived as:

$$d\boldsymbol{\sigma}^e = \mathbf{C}^e \, d\boldsymbol{\varepsilon}^e + \mathbf{h}^T \, d\bar{\boldsymbol{\varepsilon}} + \boldsymbol{\kappa}^q \, d\mathbf{U}^q + \mathbf{g}^{em} \, d\mathbf{m}^a, \tag{14}$$

$$d\bar{\boldsymbol{\sigma}} = \mathbf{C} \, d\bar{\boldsymbol{\varepsilon}} + \mathbf{h} \, d\boldsymbol{\varepsilon}^e + \mathbf{g}^m \, d\mathbf{m}^a, \tag{15}$$

$$d\mathbf{b}^a = \boldsymbol{\mu}^a \, d\mathbf{m}^a + (\mathbf{g}^m)^T \, d\bar{\boldsymbol{\varepsilon}} + (\boldsymbol{\xi}^m)^T d\mathbf{U}^q + (\mathbf{g}^{em})^T d\boldsymbol{\varepsilon}^e, \tag{16}$$

$$d\mathbf{F}^q = \mathbf{C}^q \, d\mathbf{U}^q + (\boldsymbol{\kappa}^q)^T d\boldsymbol{\varepsilon}^e + \boldsymbol{\xi}^q \, d\mathbf{m}^a, \tag{17}$$

where

$$\boldsymbol{\sigma}^e = [\sigma_{11}^e, \sigma_{22}^e, \sigma_{12}^e, \mathfrak{T}^e]^T, \tag{18}$$

$$\bar{\boldsymbol{\sigma}} = [\bar{\sigma}_{11}, \bar{\sigma}_{22}, \bar{\sigma}_{33}, \bar{\tau}_{23}, \bar{\tau}_{13}, \bar{\tau}_{12},]^T, \tag{19}$$

$$\mathbf{b}^a = [b_1, b_2, b_3, d_{111}^{re}, d_{121}^{re}, d_{131}^{re}, d_{112}^{re}, \ldots, d_{133}^{re}, d_{111}^{im}, \ldots, d_{133}^{im}, d_{211}^{re}, \ldots]^T, \tag{20}$$

$$\mathbf{F}^q = \frac{1}{2}\bigl[-\sigma_{1l}^{q_{11}s} q_l, -\sigma_{2l}^{q_{11}s} q_l, -\sigma_{3l}^{q_{11}s} q_l, \sigma_{1l}^{q_{11}c} q_l, \sigma_{2l}^{q_{11}c} q_l, \sigma_{3l}^{q_{11}c} q_l, -\sigma_{1l}^{q_{12}s} q_l, -\sigma_{2l}^{q_{12}s} q_l, \ldots,$$
$$\sigma_{3l}^{q_{12}c} q_l, -\sigma_{1l}^{q_{13}s} q_l, \ldots, \sigma_{3l}^{q_{13}c} q_l, -\sigma_{1l}^{q_{21}s} q_l, \ldots, \sigma_{3l}^{q_{23}c} q_l, \ldots\bigr]^T, \tag{21}$$

and

$$C_{ij} = \left(\frac{\partial^2 \bar{w}}{\partial \bar{\varepsilon}_i \partial \bar{\varepsilon}_j}\right)_0, \quad h_{ij} = \left(\frac{\partial^2 \bar{w}}{\partial \bar{\varepsilon}_i \partial \varepsilon_j^e}\right)_0, \quad g_{ij}^m = \left(\frac{\partial^2 \bar{w}}{\partial \bar{\varepsilon}_i \partial m_j^a}\right)_0,$$

$$C_{ij}^e = \left(\frac{\partial^2 \bar{w}}{\partial \varepsilon_i^e \partial \varepsilon_j^e}\right)_0, \quad \kappa_{ij}^q = \left(\frac{\partial^2 \bar{w}}{\partial \varepsilon_i^e \partial U_j^q}\right)_0, \quad g_{ij}^{em} = \left(\frac{\partial^2 \bar{w}}{\partial \varepsilon_i^e \partial m_j^a}\right)_0, \tag{22}$$

$$\mu_{ij}^a = \left(\frac{\partial^2 \bar{w}}{\partial m_i^a \partial m_j^a}\right)_0, \quad \xi_{ij}^m = \left(\frac{\partial^2 \bar{w}}{\partial m_i^a \partial U_j^q}\right)_0, \quad C_{ij}^q = \left(\frac{\partial^2 \bar{w}}{\partial U_i^q \partial U_j^q}\right)_0.$$

In eqs. (14-17), terms with a prefix "$d$" denote a small disturbance. In eq. (22), $\bar{w} = \frac{1}{V}\int \tilde{w} \, dV$, where $V$ denotes the volume of the material, and terms with a subscript 0 take values at equilibrium state. Eqs. (14-17) provide the formula to quantify the emergent elastic deformation of skX under the five kinds of effective external fields. In eq. (22), $\mathbf{C}^e$ denotes the second order derivatives of the free energy density with respect to the emergent elastic strains, for which it is termed the ***emergent elastic stiffness matrix*** of the skX. $\mathbf{h}$ describes the linear coupling between the emergent elastic strains $\boldsymbol{\varepsilon}^e$ and the elastic strains $\bar{\boldsymbol{\varepsilon}}$, for which it is termed the ***emergent piezoelastic matrix*** of the skX, imitating the nomenclature of piezoelectricity[28]. The analytical expressions of the components of matrices defined in eq. (22) in terms of the thermodynamic parameters are generally lengthy and thus provided in the supplementary information

Consider now a more specific case where the equilibrium state is only disturbed by homogeneous mechanical forces, we have $d\boldsymbol{\sigma}^e = \mathbf{0}$, $d\mathbf{b}^a = \mathbf{0}$, $d\mathbf{b}^q = \mathbf{0}$ and $d\mathbf{F}^q = \mathbf{0}$. After manipulation, we have from eqs. (14-17)

$$d\boldsymbol{\varepsilon}^e = \boldsymbol{\lambda} \, d\bar{\boldsymbol{\varepsilon}}, \tag{23}$$

and

$$d\boldsymbol{\varepsilon}^e = \boldsymbol{\lambda}(\mathbf{C}^* + \mathbf{h}^*\boldsymbol{\lambda})^{-1} \, d\bar{\boldsymbol{\sigma}}, \tag{24}$$

where

$$\boldsymbol{\lambda} = -(\mathbf{C}^{e**})^{-1}(\mathbf{h}^*)^T, \tag{25}$$

and

$$\begin{aligned}
\mathbf{C}^* &= \mathbf{C} - \mathbf{g}^m(\boldsymbol{\mu}^{a*})^{-1}(\mathbf{g}^m)^T \\
\mathbf{h}^* &= \mathbf{h} - \mathbf{g}^m(\boldsymbol{\mu}^{a*})^{-1}(\mathbf{g}^{em*})^T, \\
\mathbf{C}^{e**} &= \mathbf{C}^{e*} - \mathbf{g}^{em*}(\boldsymbol{\mu}^{a*})^{-1}(\mathbf{g}^{em*})^T, \\
\mathbf{C}^{e*} &= \mathbf{C}^e - \boldsymbol{\kappa}^q(\mathbf{C}^q)^{-1}(\boldsymbol{\kappa}^q)^T, \\
\mathbf{g}^{em*} &= \mathbf{g}^{em} - \boldsymbol{\kappa}^q(\mathbf{C}^q)^{-1}\boldsymbol{\xi}^m, \\
\boldsymbol{\mu}^{a*} &= \boldsymbol{\mu}^a - (\boldsymbol{\xi}^m)^T(\mathbf{C}^q)^{-1}\boldsymbol{\xi}^m.
\end{aligned} \qquad (26)$$

We call the $4 \times 6$ matrix $\boldsymbol{\lambda}$ defined in eq. (25) the ***emergent strain ratio* (ESR)** matrix. It determines the ratio between the emergent elastic strains and the elastic strains of the underlying material. In eq. (26), we see that $\boldsymbol{\lambda}$ is determined by $\mathbf{C}^{e**}$ and $\mathbf{h}^*$, where each superscript "*" denotes one renormalization of the parameter matrix due to multi-field coupling described by eqs. (14-17).

Substituting eq. (1) into eqs. (22-26) with the Fourier representation of skX, after manipulation one can derive the analytical expression in terms of the thermodynamic parameters of the underlying material, which is very lengthy and hence not presented here. An important question to ask is which thermodynamics parameters determine the emergent elastic property of the skX. To answer this question, we decompose the ESR matrix as follow

$$\boldsymbol{\lambda} = \boldsymbol{\lambda}^a + \boldsymbol{\lambda}^b, \qquad (27)$$

where $\boldsymbol{\lambda}^a = -(\mathbf{C}^{e**})^{-1}(\mathbf{h})^T$ is called the direct ESR matrix since it describes the direct coupling between $\boldsymbol{\varepsilon}^e$ and $\bar{\boldsymbol{\varepsilon}}$. $\boldsymbol{\lambda}^b = -(\mathbf{C}^{e**})^{-1}[(\mathbf{h}^*)^T - (\mathbf{h})^T]$ is called the indirect ESR matrix. It exists because $\mathbf{m}^a$ is coupled with both $\boldsymbol{\varepsilon}^e$ and $\bar{\boldsymbol{\varepsilon}}$, and so it causes indirect coupling between the latter two vectors. According to the definition of $h_{ij}$ given in eq. (22), it derives from the coupling term between $\boldsymbol{\varepsilon}^e$ and $\bar{\boldsymbol{\varepsilon}}$ in the free energy density functional, i.e., $\sum_{i=1}^{6} \tilde{L}_{Oi} \tilde{f}_{Oi}$ introduced in eq. (3). After deduction, we have approximately $\lambda_{ij}^a \propto \tilde{L}_{O1}, \tilde{L}_{O2}, \tilde{L}_{O3}$. On the other hand, $\boldsymbol{\lambda}^b = (\mathbf{C}^{e**})^{-1}[\mathbf{g}^{em*}(\boldsymbol{\mu}^{a*})^{-1}(\mathbf{g}^m)^T]$ corresponds to a very complex expression. Nevertheless, we find that in most conditions of temperature and magnetic field $\lambda_{ij}^b \propto \tilde{K}$. Since we have for the magnetoelastic coupling coefficients $\tilde{K} \gg \tilde{L}_i \gg \tilde{L}_{Oi}$, the indirect ESR matrix $\boldsymbol{\lambda}^b$ dominates the emergent elastic property of skX in most cases, which is two orders of magnitude larger than $\boldsymbol{\lambda}^a$. This has been validated by our calculation of $\lambda_{ij}$ for MnSi in the next section. Revisiting the definition of the rescaled magnetoelastic coefficients[21], we have $\tilde{K} = \frac{4AK}{D^2 M_s^2}$, where $K$ is the magnetoelastic coefficient, $A$ is the exchange interaction stiffness, $D$ is the Dzyaloshinskii-Moriya (DM) coupling coefficient, and $M_s$ denotes the saturation magnetization examined near absolute zero. Since the deformability of skX under mechanical loads is linearly proportional to $\tilde{K}$, the upper definition provides us a clue to search for materials hosting skX that is more sensitive to mechanical loads.

## III. Emergent elastic property of the skX in MnSi

Following the procedure introduced in the method section, we analyze the emergent elastic property of the skX in a prototype helimagnet MnSi by calculating variation of $\lambda_{ij}$ in the temperature-magnetic field phase diagram, where we assume that bias magnetic field satisfies $\mathbf{b} = [0, 0, b]^T$.

We find that **a)** all the components of $\boldsymbol{\lambda}$ are very sensitive to variation of the magnetic field. More specifically, the dominant $\lambda_{ij}$ reverses its sign as the magnetic field increases (Figure 2). It means when concerning the deformation of skX caused by a certain form of mechanical loading under two different conditions of magnetic field, one may observe two emergent deformation patterns that are opposite to each other, as shown in Figure 1(i-p). To understand this phenomenon, we consider the definition of $\boldsymbol{\lambda}$, which reflects the sensitivity of $d\boldsymbol{\varepsilon}^e$ to $d\bar{\boldsymbol{\varepsilon}}$. It is mentioned above that the emergent elastic property of skX is dominated by $\widetilde{K}$, which appears through the magnetoelastic coupling term $\widetilde{K}\mathbf{m}^2\varepsilon_{ii}$ in eq. (3). Considering this dominant magnetoelastic coupling term only, a small variation of elastic strain $d\varepsilon_{11}$ (or $d\varepsilon_{22}$, $d\varepsilon_{33}$) is equivalent to a renormalization of the second order Landau expansion term $t\mathbf{m}^2$ in eq. (1), which turns to $(t + \widetilde{K} d\varepsilon_{11})\mathbf{m}^2$. For a negative $\widetilde{K}$, as is the case for MnSi, a positive $d\varepsilon_{11}$ is equivalent to a decrease of the rescaled temperature $t$. Therefore, the variation of $\lambda_{i1}, \lambda_{i2}, \lambda_{i3}$ with $b$ should behave similarly to the variation of $-\frac{(d\varepsilon^e)_i}{dt}$ with $b$. For MnSi, this is validated by comparing Figure 3(a) with Figure 3(e), where the variation of $\lambda_{11}, \lambda_{12}, \lambda_{13}$ and $-\frac{d\varepsilon_{11}^e}{dt}$ with $b$ are plotted respectively.

**b)** The dominant $\lambda_{ij}$s ($\lambda_{11}, \lambda_{12}, \lambda_{13}, \lambda_{21}, \lambda_{22}, \lambda_{23}$) diverge near the critical point of a skX-ferromagnetic phase transition in the temperature-magnetic field phase diagram (Figure 3a). The origin of this divergence is two-fold. Through soft-mode analysis[20], we know that the skX-ferromagnetic phase transition is accompanied by a vanishing of both $\mathbf{m}_{\mathbf{q}_{ij}}$ (hence a singular $\boldsymbol{\mu}^{\mathbf{a}*}$) and the skyrmion-skyrmion interaction (hence a singular $\mathbf{C}^{e**}$, illustrated in Figure 3d). According to eqs. (25, 26), a singular $\boldsymbol{\mu}^{\mathbf{a}*}$ leads to a divergent $\mathbf{h}^*$ (Figure 3c), and a singular $\mathbf{C}^{e**}$ leads to a divergent $(\mathbf{C}^{e**})^{-1}$. The divergence of $\lambda_{ij}$ during a skX-ferromagnetic phase transition is thus anticipated.

**c)** The values of all components of $\boldsymbol{\lambda}$ are very sensitive to the order of Fourier representation, as illustrated in Figure 3b-d. Similar to the variation of lattice constant of skX with $b$[20], we find that $3^{\text{rd}}$ order Fourier representation is necessary to provide quantitatively reliable values of $\lambda_{ij}$. As explained before[20], the dependence of $\lambda_{ij}$ on the Fourier representation order stems from the coupling between Fourier terms with different order permitted by the $\mathbf{m}^4$ term in eq. (1).

**d)** $\lambda_{ij}$ out-of-plane shear strains $\varepsilon_{13}$ and $\varepsilon_{23}$ do not induce any emergent deformation of skX, and the sensitivity of emergent rotation $\omega^e$ to mechanical loads is at least an order of magnitude smaller than the other emergent elastic strains.

## IV. Conclusion

We derive one of the basic equations, the linear constitutive equations, in emergent elasticity of skX. This set of equations study the deformation of skX under application of external forces and other effective fields. Based on the equations, we not only clarify that the emergent elastic property of skX stems from the magnetoelastic coupling of the underlying chiral magnets, but also establish the analytical relations between the components of the ESR matrix $\boldsymbol{\lambda}$ and the magnetoelastic coefficients in the thermodynamics model. Calculation of $\lambda_{ij}$ for MnSi shows that the dominant $\lambda_{ij}$s ($\lambda_{11}, \lambda_{12}, \lambda_{13}, \lambda_{21}, \lambda_{22}, \lambda_{23}$) change sign as magnetic field increases, which means that we can obtain two opposite deformation patterns of skX under the same applied mechanical loads by varying the bias magnetic field. Since MnSi has only moderate magnetoelastic coupling strength among all ferromagnets, and yet the dominant $\lambda_{ij}$s can be as large as 20 at low bias field where skX is stable, our results imply that emergent crystals possess novel macroscopic properties that are generally sensitive to the application of mechanical forces. The equations and methodology developed in this work is of fundamental significance for the development of emergent elasticity, a new research area with promising future.

## Methods

### a) Details of the Fourier representation of skX

Substituting the nth order Fourier representation $\mathbf{m}_{Fn} = \mathbf{m_0} + \sum_{i=1}^{n}\sum_{j=1}^{n_i} \mathbf{m}_{\mathbf{q}_{ij}} e^{i\mathbf{q}_{ij}\cdot\mathbf{r}}$ of skX into eq. (1), after integration we have

$$\bar{w}(\mathbf{m}_{Fn}) = -\frac{(t-1)^2}{4} - b^2 + (\mathbf{m_0} - \mathbf{b})^2 + \bar{w}_{int} + \bar{w}_{per}, \tag{28}$$

where $\bar{w}(\mathbf{m}_{Fn}) = \frac{1}{V}\int \tilde{w}(\mathbf{m}_{Fn})dV$, $\bar{w}_{int} = \frac{1}{V}\int \left(\mathbf{m}_{Fn}^2 + \frac{t-1}{2}\right)^2 dV$ includes the effect of nonharmonic mode-mode interaction, and $\bar{w}_{per} = \sum_{i=1}^{\infty}\sum_{j=1}^{n_i} \left(\mathbf{m}_{\mathbf{q}_{ij}}^*\right)^T \mathbf{A}_{ij} \mathbf{m}_{\mathbf{q}_{ij}}$ includes the effect of all the gradient terms. Here $\mathbf{m}_{\mathbf{q}_{ij}}^*$ denotes the conjugate of $\mathbf{m}_{\mathbf{q}_{ij}}$, and

$$\mathbf{A}_{ij} = \begin{bmatrix} 1+(s_iq)^2 & 0 & 2iq_{ijy} \\ 0 & 1+(s_iq)^2 & -2iq_{ijx} \\ -2iq_{ijy} & 2iq_{ijx} & 1+(s_iq)^2 \end{bmatrix}, \tag{29}$$

where $|\mathbf{q}_{ij}| = s_{ij}q = \sqrt{q_{ijx}^2 + q_{ijy}^2}$ and $i = \sqrt{-1}$. The eigenvalues of $\mathbf{A}_{ij}$ are $(s_iq - 1)^2$, $(s_iq)^2$, and $(s_iq + 1)^2$, and the corresponding unit eigenvectors can be solved as $\mathbf{P}_{ij1} = \frac{1}{\sqrt{2}s_iq}[-iq_{ijy}, iq_{ijx}, s_iq]^T$, $\mathbf{P}_{ij2} = \frac{1}{s_iq}[q_{ijx}, q_{ijy}, 0]^T$, $\mathbf{P}_{ij3} = \frac{1}{\sqrt{2}s_iq}[iq_{ijy}, -iq_{ijx}, s_iq]^T$. Hence $\mathbf{m}_{\mathbf{q}_{ij}}$ can be expanded as a linear combination of the three unit eigenvectors

$$\mathbf{m}_{\mathbf{q}_{ij}} = c_{ij1}\mathbf{P}_{ij1} + c_{ij2}\mathbf{P}_{ij2} + c_{ij3}\mathbf{P}_{ij3}, \tag{30}$$

and

$$c_{ijk} = c_{ijk}^{re} + ic_{ijk}^{im}, \tag{31}$$

where $c_{ijk}^{re}$ and $c_{ijk}^{im}$ are real variables to be determined. The above deduction explains the definition of $\mathbf{m}^{\mathbf{a}}$ introduced in eq. (13). On the other hand, the work conjugates of $c_{ijk}^{re}$ and $c_{ijk}^{im}$ are defined in eq. (20) as $d_{ijk}^{re}$ and $d_{ijk}^{im}$, which are related to $\mathbf{b}_{q_{ij}}^*$, amplitude of a spatially periodic magnetic field $\mathbf{b}_{q_{ij}}^* e^{-i\mathbf{q}_{ij}\cdot\mathbf{r}}$ by

$$\mathbf{b}_{q_{ij}}^* = d_{ij1}\mathbf{P}_{ij1}^* + d_{ij2}\mathbf{P}_{ij2}^* + d_{ij3}\mathbf{P}_{ij3}^*, \tag{32}$$

and

$$d_{ijk} = \frac{1}{2}\left(d_{ijk}^{re} - id_{ijk}^{im}\right). \tag{33}$$

Here $\mathbf{P}_{ijk}^*$ denotes the complex conjugate of $\mathbf{P}_{ijk}$.

### b) The standard procedure to calculate $\lambda_{ij}$ at given temperature-magnetic field

At rescaled temperature $t$ and rescaled magnetic field $b$, minimize the averaged free energy density $\bar{w} = \frac{1}{V}\int \tilde{w}\, dV$ with respect to all the independent variables introduced in eqs. (10-13), which determines their equilibrium values at given $t$ and $b$. One should notice that when the system is free from other effective fields, we will have approximately $(\varepsilon_{12}^e)_0 = (\omega^e)_0 = 0$, and $(\varepsilon_{11}^e)_0 = (\varepsilon_{22}^e)_0 = \frac{1}{(q)_0} - 1$ where $q = 1$ is assumed in the minimization process. This relation tells us how to solve $(q)_0$, which describes the actual lattice constant of skX at given $t$ and $b$.

The equilibrium value of independent variables is used to calculate the compliance matrices defined in eq. (22) at given $t$ and $b$, where we should use $q = (q)_0$ and $(\varepsilon_{11}^e)_0 = (\varepsilon_{22}^e)_0 = (\varepsilon_{12}^e)_0 = (\omega^e)_0 = 0$. Then we can calculate $\lambda_{ij}$ through eqs. (25, 26).

Table 1. $\lambda_{ij}$ for bulk MnSi calculated at two conditions: (a) $t = 0.5, b = 0.1$, and (b) $t = 0.5, b = 0.3$.

| $\lambda_{ij}$ | $\bar{\varepsilon}_{11}$ | $\bar{\varepsilon}_{22}$ | $\bar{\varepsilon}_{33}$ | $\bar{\gamma}_{23}$ | $\bar{\gamma}_{13}$ | $\bar{\gamma}_{12}$ |
|---|---|---|---|---|---|---|
| $\varepsilon_{11}^e$ | (a) 14.70 | (a) 5.50 | (a) 13.47 | (a) 0 | (a) 0 | (a) -0.023 |
| | (b) -17.69 | (b) -21.16 | (b) -18.45 | (b) 0 | (b) 0 | (b) -0.0087 |
| $\varepsilon_{22}^e$ | (a) 7.73 | (a) 17.31 | (a) 8.78 | (a) 0 | (a) 0 | (a) 0.045 |
| | (b) -19.90 | (b) -17.08 | (b) -19.53 | (b) 0 | (b) 0 | (b) -0.0027 |
| $\varepsilon_{12}^e$ | (a) -0.11 | (a) -0.11 | (a) 0.088 | (a) 0 | (a) 0 | (a) -3.20 |
| | (b) -0.23 | (b) -0.23 | (b) -0.21 | (b) 0 | (b) 0 | (b) -0.92 |
| $\omega^e$ | (a) -0.63 | (a) 1.12 | (a) -0.48 | (a) 0 | (a) 0 | (a) -0.38 |
| | (b) -0.29 | (b) 0.50 | (b) 0.21 | (b) 0 | (b) 0 | (b) -0.17 |

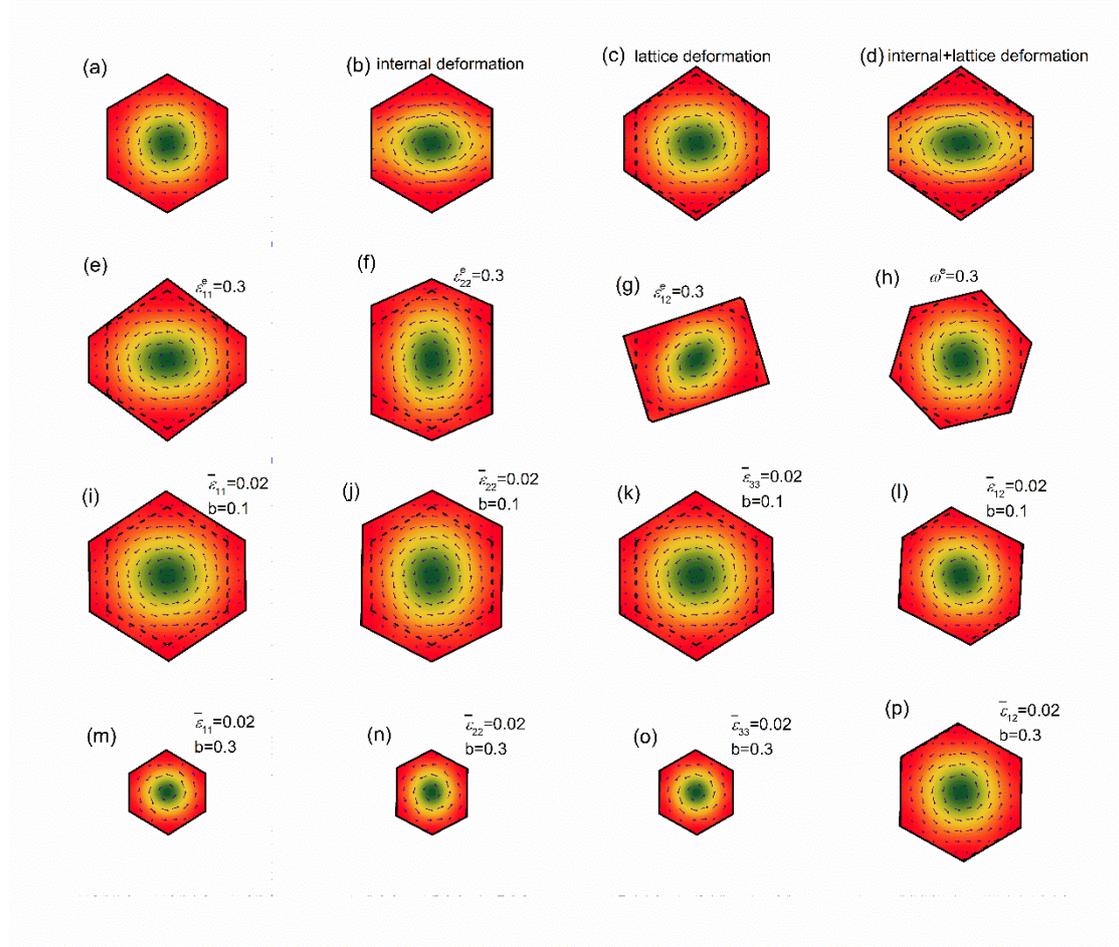

Figure 1. Wigner-Seitz cell of the skX under various conditions: (a) undeformed, (b) with internal deformation only, (c) with lattice deformation only, (d) with both internal and lattice deformation. (e-h) present the cells after four basic modes of emergent deformation: (e) $\varepsilon_{11}^e = 0.3$, (f) $\varepsilon_{22}^e = 0.3$, (g) $\varepsilon_{12}^e = 0.3$, (h) $\omega^e = 0.3$. (i-p) plot the deformed Wigner-Seitz cell of the skX when the four

types of effective elastic strains are applied, where (i-l) are calculated at the condition $t = 0.5, b = 0.1$, and (m-p) are calculated at the condition $t = 0.5, b = 0.3$: (i and m) $\bar{\varepsilon}_{11} = 0.02$, (j and n) $\bar{\varepsilon}_{22} = 0.02$, (k and o) $\bar{\varepsilon}_{33} = 0.02$, (l and p) $\bar{\varepsilon}_{12} = 0.02$. The vectors illustrate the distribution of the in-plane magnetization components with length proportional to their magnitude, while the colored density plot illustrates the distribution of the out-of-plane magnetization component. The black dashed line plots the undeformed Wigner-Seitz cell, while the black solid line plots the deformed cell.

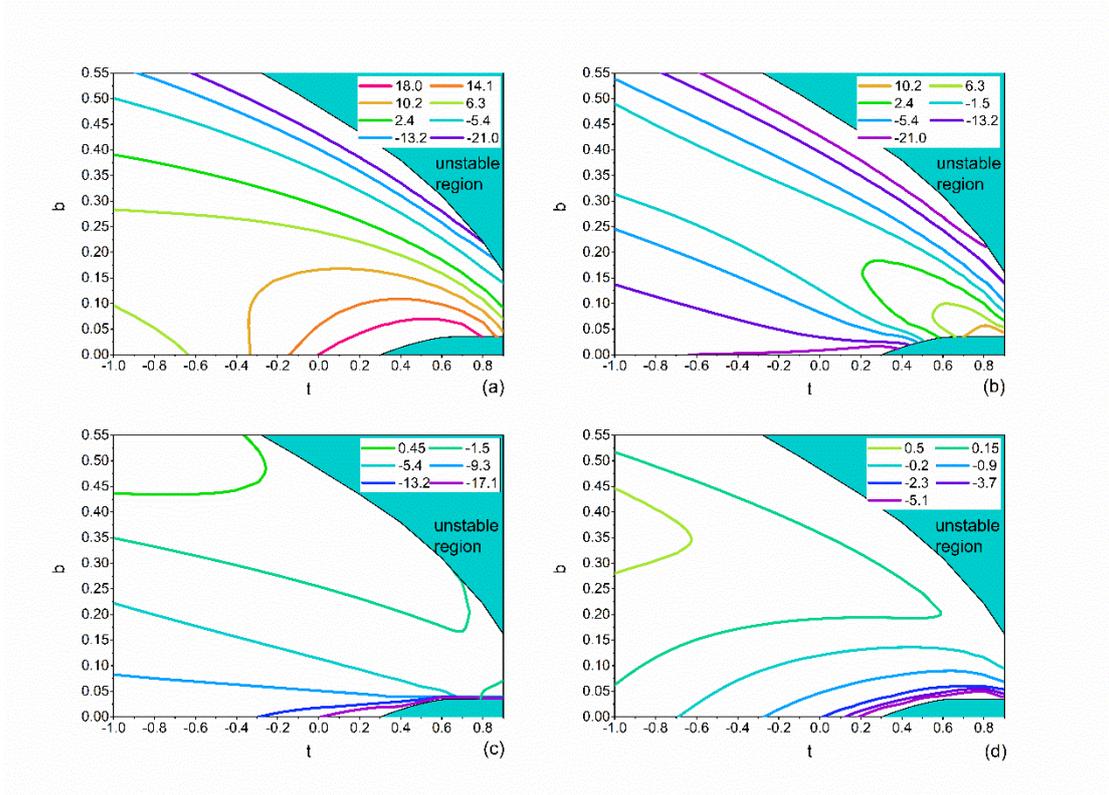

Figure 2. Contour plot of components of the ESR matrix for bulk MnSi in the $t - b$ phase diagram: (a) $\lambda_{11}$, (b) $\lambda_{12}$, (c) $\lambda_{36}$, (d) $\lambda_{41}$. The cyan region illustrates the area where the skX becomes intrinsically unstable. The distribution of $\lambda_{13}$, $\lambda_{23}$, and $\lambda_{22}$ in the $t - b$ phase diagram resembles that of $\lambda_{11}$; the distribution of $\lambda_{21}$ resembles that of $\lambda_{12}$; $\lambda_{31}, \lambda_{32}$ and $\lambda_{33}$ are generally negligible compared with $\lambda_{36}$ and hence not of interest; and the distribution of $\lambda_{42}, \lambda_{43}$ and $\lambda_{46}$ resembles that of $\lambda_{41}$.

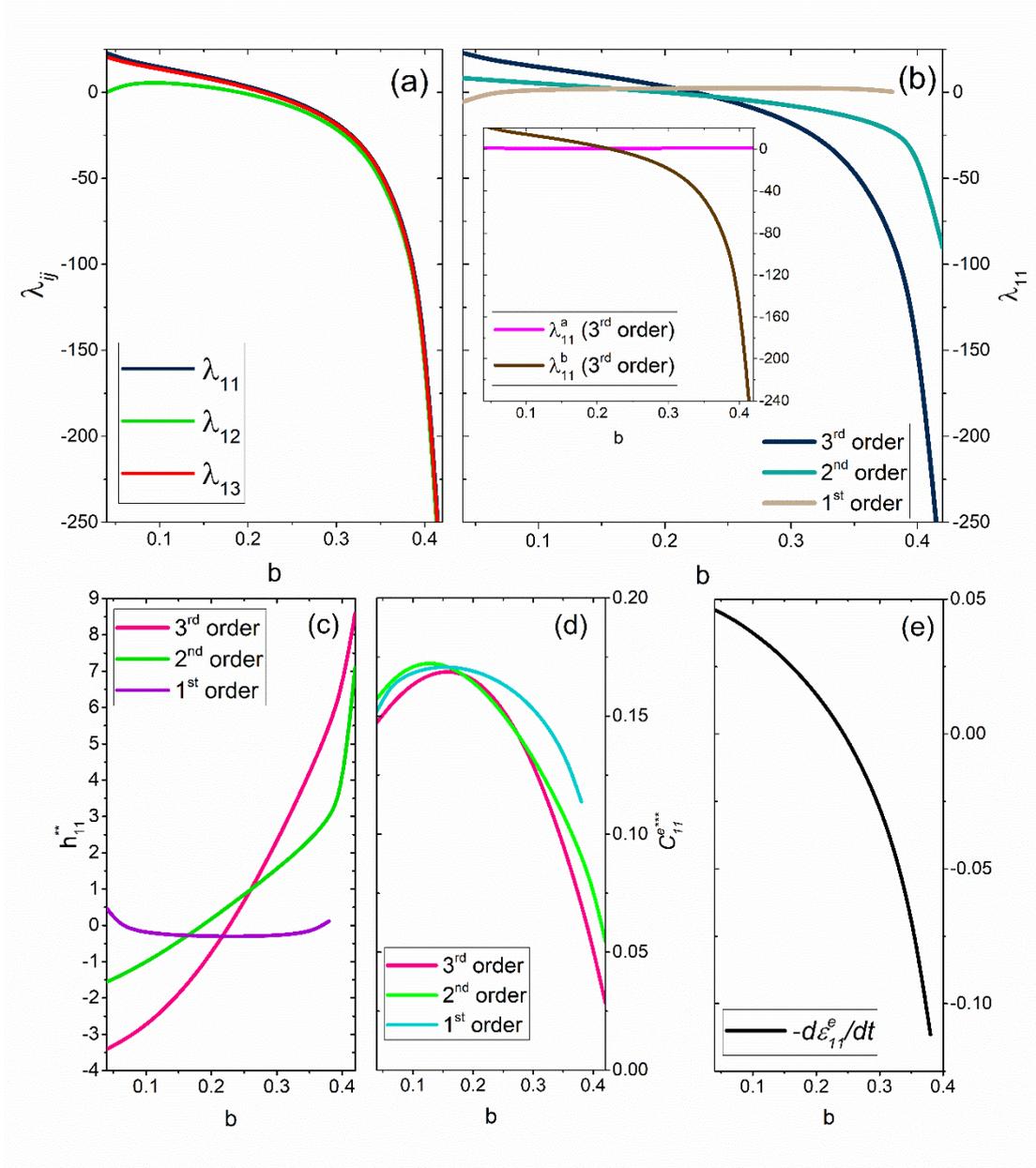

Figure 3. Variation of (a) $\lambda_{1j}, (j = 1, 2, 3)$, (b) $\lambda_{11}$, (c) $h_{11}^{**}$, and (d) $C_{ii}^{e***}$ obtained through different orders of Fourier representation, (e) $-\frac{d\varepsilon_{11}^e}{dt}$ with the rescaled magnetic field $b$ calculated for bulk MnSi at the condition $t = 0.5$. The inset of (b) plots the variation of $\lambda_{11}^a$ and $\lambda_{11}^b$ with the magnetic field.